\documentstyle[12pt]{article}
\def\Journal#1#2#3#4{{#1}{\bf #2}, #3 (#4)}

\def\NPB{{\em Nucl. Phys.} B}
\def\PLB{{\em Phys. Lett.} B}
\def\PRL{\em Phys. Rev. Lett.~}
\def\PRD{{\em Phys. Rev.} D}

\def\be{\begin{equation}}
\def\ee{\end{equation}}
\def\bea{\begin{eqnarray}}
\def\eea{\end{eqnarray}}

\newcommand{\beq}{\begin{equation}}
\newcommand{\eeq}{\end{equation}}
\newcommand{\EQ}{\begin{equation}}
\newcommand{\EN}{\end{equation}}

\def\ne{\hbox{$\nu_e$ }}

\def\nm{\hbox{$\nu_\mu$ }}

\def\nt{\hbox{$\nu_{\tau}$ }}

\begin{document}

\rightline{IC/96/242}

\begin{center}

{\LARGE \bf Solar Neutrino Spectroscopy\\ 
(Before and After SuperKamiokande)\footnote{Talk 
given at XVII International conference on 
{\it Neutrino Physics and Astrophysics}, Helsinki, Finland, 
June 13-19, 1996.}\\
}
\vskip 1truecm

{ A. Yu. Smirnov}

{\it International Centre for Theoretical Physics\\
Strada Costiera 11, 34100 Trieste, Italy,\\
Institute for Nuclear Research, Russian Academy of Sciences,\\
117312 Moscow, Russia
}

\vskip 3truecm

{\bf Abstract}
\end{center}

Results of solar neutrino spectroscopy based on data from 
four experiments are presented. 
Perspectives related to forthcoming experiments are discussed. 
Implications 
of the results for  neutrino properties are considered.\\ 


\newpage

\section{Introduction}

More than 30 years ago Bahcall~\cite{ba}  
and independently Zatsepin and Kuzmin~\cite{zat}    
had put forward 
the  program  of the complete solar neutrino spectroscopy. 
The idea was to perform several (radiochemical) 
experiments with different 
absorbers and absorption thresholds
which are sensitive to different parts of the 
$\nu_{\odot}$-spectrum. This 
allows one to find   fluxes of
different components of the spectrum. 
In turn, comparing the fluxes one can 
study both  interior of the Sun and 
(as it was realized later) properties of neutrinos themselves. 
Along with this line the Gallium 
experiment was proposed~\cite{kuz}. 
That time Reines~\cite{rei} started to 
use direct electronic methods to get the information about 
$\nu_{\odot}$
-fluxes.  

In eighties, an  analysis of the  Homestake data had risen 
the question about time variations of 
$\nu_{\odot}$ - fluxes. 
Searches for possible 
time variations give  
very important and complementary  
information both about physics 
of the Sun and on neutrinos. 

Great success of  helioseismology in studies of  
interior of the Sun shifted the interests  
to the second aspect of solar neutrino spectroscopy: properties 
of neutrinos.

Today there are the  data from five experiments: 
Homestake~\cite{hom}, Kamiokande~\cite{kam}, SAGE \cite{sag}, 
 GALLEX~\cite{gal} and first preliminary results from SuperKamiokande 
\cite{SK}. 
Already with these data we can perform tentative spectroscopy 
of solar neutrinos. Forthcoming and future experiments 
will open a possibility to realize the program completely.

\section{Neutrino Spectroscopy}

\subsection{Original fluxes and conversion probabilities}

Fluxes of the electron neutrinos  in  the Earth detectors,  
$F_i(E)$, ($i = pp, Be, pep$, $N$, $O, B $), can be written as 
\be
F_i (E)  = P(E)  \cdot f_i \cdot  F_i^{0}(E)~,  
\label{flux}
\ee
where $F_i^0$ are the fluxes in the reference standard solar model 
(RSSM).  
And in what follows we will use the model \cite{bp96} as the reference one.

Factors $P(E)$  are the electron neutrino survival probabilities:  
$\ne \rightarrow \ne$. They describe possible 
effects of  neutrino transformations   
and  can be called the {\it neutrino 
factors}. These factors depend on neutrino parameters: 
$\Delta m^2, \theta, E, \mu  $ as well as on characteristics of the 
Sun -- density, magnetic fields  {\it etc.},  
and satisfy the restriction
$0 \leq P(E, \theta,  \Delta m^2) \leq 1$.

On the contrary, $f_i$ are the {\it solar model factors} which 
describe deviation of  true original neutrino fluxes  
from those predicted by the reference SSM. 
The product $f_i \cdot F^0_i$ is the original flux of $i$-component,   
so that  $f_i$ can be considered as  the flux in the  units 
of the reference model flux.\\  

\noindent
{\it The problem} is to find separately $f_i$ and 
$P(E)$ from the solar neutrino data. 
Although in (\ref{flux}) they enter as the product,    
there are two 
features which allow one to distinguish their effects. 
\begin{itemize}

\item  In general, $P$ depends on neutrino energy, 
whereas $f_i =const$; 

\item In the case of the flavor conversion 
the fluxes of \nm and \nt appear. 
If there is no transition  to sterile 
neutrinos\footnote{If there is the transition  to 
both active and sterile neutrinos, the   
flux of \nm, \nt is proportional to   $(1 - P - P_s)$, where 
$P_s$ is the probability of $\nu_e \rightarrow S$.
},     
 these fluxes equal 
\be
F_i (\nm, \nt) = (1 - P)~ f_i~ F_i^{0}~.   
\label{muflux}
\ee
They depend  not only on 
the product $f\cdot P$, as in (\ref{flux}),  
but also separately on $f_i$. 
Neutrinos \nm and \nt  contribute to signals  
(e.g.  $\nu - e$ scattering, $\nu d \rightarrow \nu n p$)  
due to  neutral current interactions. 

\end{itemize} 
 
If the Sun is ``standard",  then  $f_i = 1$. 
For ``standard" neutrinos (massless, unmixed) we take  
$P = 1$.

\subsection{Boron neutrinos: implications of Kamiokande and 
SuperKamiokande results}

It is worthwhile  to start the analysis by  the Kamiokande data 
for two reasons: 
(i) The data are  sensitive to one component of the spectrum -- 
boron neutrinos only \footnote{The contribution from Hep neutrinos 
is negligibly small.},  and therefore  
have simple interpretation. (ii) 
In near future SK and SNO \cite{sno} will perform precise measurements of 
the boron 
neutrino flux, and this flux  will be the reference point in the spectrum. 

Basic conclusions one can draw from the Kamiokande result are the 
following. 

\begin{itemize} 

\item
Kamiokande has certainly detected the Boron neutrinos: 
The shape of energy spectrum 
of the recoil electrons 
corresponds to  original spectrum of the boron neutrinos. 
In particular, maximal energies of the electrons 
correspond to the end 
point of neutrinos from  the $^8 B$ - decay.  

\item
The boron 
neutrinos are produced in the reaction at the end of the 
$pp$-III - nuclear cycle. Therefore the Kamiokande result means  that 
all previous reactions which lead to production of 
$^8 B$ take place,  {\i.e.}   the  complete  
$pp$-III -cycle operates inside the Sun.

\item
Once the  $pp$-cycle operates,  we know 
the components of the neutrino spectrum as well as  
energy dependencies of these individual components
which are determined essentially by 
known kinematics of the weak interaction processes. 

\end{itemize} 

However all this  does not allow one to restore 
the absolute values of fluxes. 
Moreover, present solar neutrino data  do not even prove that 
$pp$-cycle is the main source of energy of the Sun. 
In \cite{bfk} it was shown that neutrino 
data themselves do not contradict hypothetical possibility 
that $CNO$-cycle dominates in the  energy release. 
The  gallium  data can be  explained by fluxes of   
$N-$ and $O-$ neutrinos. An  agreement with observations 
can be achieved,   if one  suggests that there is a conversion of 
neutrinos ({\it e.g.} the resonance conversion) which 
strongly suppresses  fluxes in the energy range 0.8 - 2 MeV.\\ 

\noindent
{\it Spectroscopy with Kamiokande.}  
Kamiokande data (although being in agreement with undistorted 
neutrino spectrum) do not exclude an   appreciable distortion 
of spectrum.  
Let us introduce the ratio of the observed energy spectrum 
of the recoil electrons $N_e$ and  the one expected from 
the ``standard" boron neutrinos, $N_e^{SSM}$: 
\be
R_e (E) \equiv  \frac{N_e (E)}{N_e^{SSM} (E)}~. 
\ee
Obviously, $R_e(E) = 1$ in absence of distortion. 
Due to strong smoothing in the recoil electron spectrum, 
even strong distortion of the 
neutrino spectrum leads to 
approximately linear energy dependence  of the ratio: 
\be
R_e (E) = R_e (E_0) [ 1 + s_e \cdot (E - E_0)]~. 
\label{distor}
\ee
That is 
the  distortion can be characterized by the {\it slope   
parameter}, $s_e$, which equals   
$
s_e \equiv \left[ (d R_e/dE)/R_e \right]_{E_0}. 
$
We fix the slope in the middle of the detected interval 
at $E_0 = 10$ MeV. 
The $\chi^2$ fit  
of  Kamiokande spectrum by (\ref{distor}) gives  
\be 
s_e = 0.04 \pm 0.05~~     {\rm MeV}^{-1}, ~~~  (1\sigma)~.  
\ee
The best fit point  corresponds to $\chi^2 = 9$  for 6 d.o.f.. 

Obviously $s_e$ does not depend on $f_B$,  and 
if  nonzero,  determines parameters of 
neutrinos: $s_e = s_e(\Delta m^2, \sin^2 2\theta, ...)$.  
For a given solution of the  solar neutrino problem 
$s_e$ is the  function  of the average 
conversion probability : 
$s_e = s_e(\langle P_B \rangle)$. 
Thus measurements of the slope determine 
$\langle P_B \rangle$ independently of $f_B$: 
$$
\langle P_B \rangle = 
\langle P_B (s_e)\rangle 
$$
Once $\langle P_B \rangle$ is known,  from (\ref{flux}) one can find   
$$
f_B \approx \frac{R_e (E_0)}{\langle P_B \rangle} ~. 
$$
Therefore measurements of the slope and 
the absolute value 
of the recoil electrons counting rate allow one 
to find  separately 
the original flux and the conversion probability. 

SK will improve  the sensitivity of  
measurements of $s_e$ by one order of magnitude, so that in future 
one may get the numbers like $s_e = 0.040 \pm 0.005$ 
(if {\it e.g.} the average value of the slope will be equal to present 
value). 
The accuracy in measurements of the absolute value of the flux 
will be mainly due to systematic uncertainties.  


\subsection{Spectroscopy in  absence of 
\nm and \nt components}

Results of spectroscopy depend  crucially 
on whether  neutrinos of other flavors 
are present in the solar flux. 
In future an  admixture of the non-electronic flavor will be measured by 
the SNO-experiment. 
At present we should consider two  possibilities. 

Let us assume first that non-electron neutrino fluxes are absent or 
small. This corresponds to the  astrophysical solutions or  to conversion 
of \ne into sterile neutrino. We also assume 
(as it is implied by models of the Sun) that $pp$-cycle 
gives dominant contribution to the energy release. 

Present day analysis can be done in  three steps:\\

\noindent
{\it Step 1. Kamiokande.} 
If there is no \nm- and \nt- fluxes from the Sun,  the 
Kamiokande signal is stipulated by \ne-flux 
completely: 
\be
F_B = F^{kam}_B = (2.80 \pm 0.38)\cdot 10^{6}~~ 
{\rm cm}^2 {\rm s}^{-1}~.  
\ee 
If the distortion of the spectrum is relatively weak  
we find: 
\be
f_B \cdot 
\langle P_B \rangle 
= R_{\nu e } = 0.42 \pm 0.06 
\label{boron}
\ee
in the range $E = 7 - 14$ MeV. 
Here $\langle P_B \rangle$ is the average conversion probability.  
\\

\noindent
{\it Step 2. Kamiokande and Homestake}  
\cite{bar} - \cite{smi}. 
The contribution of the boron neutrino flux  measured by Kamiokande 
to the $Ar$ - production rate is 
\be
Q_{Ar}^B = \int dE 
\sigma_{Ar}(E) F_B (E) 
= (3.1 \pm 0.4)~ {\rm  SNU} ~. 
\label{qar}
\ee 
This contribution only weakly depends on possible distortion of spectrum. 
Indeed, it turns out that  the  $Ar$-production 
cross section and the integral of the product of 
the $\nu -e$   
differential cross section  and 
the efficiency of registration in Kamiokande have approximately the 
same energy dependence~\cite{barg}: 
\be
\sigma_{Ar}(E)   \sim 
\int dE_e dE'_e 
\frac{d \sigma_{\nu e}(E, E'_e)}{dE'_e}  
K( E_e,  E'_e)~.  
\label{relation}
\ee  
Here  $K( E_e,  E'_e)$  
is the probability that the electron with energy 
$E'_e$ is detected as having the energy 
$E_e$. For this reason  
both experiments have approximately the same sensitivity to 
the neutrino spectrum. 
 
Subtracting the contribution (\ref{qar}) from the experimental 
value of the $Ar$-production rate we get
\be
Q_{Ar}^{exp} - Q_{Ar}^B = 
Q_{Ar}^{Be} + 
Q_{Ar}^{pep} + 
Q_{Ar}^{NO} 
= - 0.6 \pm 0.5 ~ {\rm SNU} , 
\ee
where $Q_{Ar}^{i}$ ($i = Be, pep, N, O$) are the contributions of 
different  components to the $Ar$-production rate.  
Putting $Q_{Ar}^{pep} + Q_{Ar}^{NO} = 0$, we get the upper bound on the 
product: 
\be
f_{Be} P_{Be} = - 0.44 \pm 0.36~.  
\label{ber}
\ee 
If $pep, N, O$ contributions equal to those in the reference 
model,  then 
$f_{Be} P_{Be} = - 1.25  \pm 0.40$.   

The SK will strengthen this result. 
(After SK the bound will be determined essentially by 
systematics of SK and the error bars of the Homestake experiment). 

The negative value of 
$f_{Be} P_{Be}$  
testifies for  presence of  neutrino flux 
which contributes to the Kamiokande signal and does not contribute to 
the Homestake signal. This can be the flux of the 
\nm and \nt (or $\bar{\nu}_{\mu}$,  $\bar{\nu}_{\tau}$) 
produced in  the conversion  \ne $\rightarrow$ \nm, \nt. 
Note that due to equality (\ref{relation}) it is impossible to 
avoid the negative value of the product in (\ref{ber}) 
by distortion of spectrum.  
Thus precise measurements of the flux 
(being confronted with Homestake Chlorine as well as Iodine 
data)  may give the proof of the neutrino flavor conversion,  
even without observations of  distortion of 
the spectrum in SK.  

Negative value of the product   disfavors also the conversion into 
sterile neutrinos. \\

\noindent
{\it Step 3. Kamiokande, Homestake and Gallium}
\cite{kwo,par,degl} .    
The contribution of the boron neutrino flux 
(as measured in Kamiokande) to the $Ge$-production rate 
is 
$Q_{Ge}^{B} = (7 \pm 1)$ SNU. For  
$Q_{Ge}^{Be} 
= Q_{Ge}^{pep} = Q_{Ge}^{N, O} = 0$, 
as it is implied by Homestake and Kamiokande results,  
we get 
\be
Q_{Ge}^{pp} = Q_{Ge}^{exp} - Q_{Ge}^{B} = 63 \pm 8~~ {\rm SNU}~ , 
\ee
and then 
\be
f_{pp} \langle P_{pp}(E) \rangle = 0.90 \pm 0.11 ~.
\label{fpp}
\ee
Here the error bars are due to  experimental errors of the GALLEX 
only.  At $2\sigma$ level non zero flux of $^7 Be$ neutrinos is 
admitted: $f_{Be} P_{Be} = 0.28$. This gives the contribution 
$Q_{Ge}^{Be} = 11$ SNU. Now the $pp$-neutrino flux should be suppressed 
stronger to satisfy the Gallium results: 
$
f_{pp} \langle P \rangle = 0.74 \pm 0.11 . 
$
If one takes into account the negative value of 
$f_{Be} P_{Be}$, then 
$
f_{pp} \langle P_{pp} \rangle = 1.14 \pm 0.31 . 
$

The results of the spectroscopy are summarized in 
the line \#1 of Table I.  
Main features are: moderate suppression at high energies 
$E > 7$ MeV; strong suppression 
at the intermediate energies and weak (or absence of) 
suppression at low energies. 
The energy independent suppression is more than 
$5\sigma$ out of the data. \\

\begin{table}[t]
\caption{ 
The spectroscopy of the solar neutrinos
\label{tab:spectrum}
}
\vspace{0.4cm}
\begin{center}
\begin{tabular}{|c|c|c|c|c|c|}
\hline 
   & E, MeV  & 0.23 - 0.42  & 0.86  & 7 - 15  &    \\
\# & $\langle P_B \rangle $  
& $f_{pp}\langle P_{pp} \rangle$ 
& $f_{Be}\langle P_{Be} \rangle$ 
& $f_{B}\langle P_{B} \rangle $ 
& $f_{B}$ \\
\hline 
1 & 1   & $1.14 \pm 0.31$ & $- 0.44 \pm 0.36$ & $0.42 \pm 0.06$ &   \\

2 & 0.5 & $0.99 \pm 0.31$ & $-0.16 \pm 0.35$  & $0.36 \pm 0.05$ & $0.73 
\pm 0.10$ \\

3 & 0.4 & $0.89 \pm 0.30$ & $- 0.03 \pm 0.33$ & $0.34 \pm 0.05$ & $0.88 \pm 
0.12$ \\

4 & 0.3 & $0.81 \pm 0.30$ & $0.16 \pm 0.31$ & $0.31 \pm 0.045$ & $1.05 \pm 
0.15$  \\

5 & 0.2 & $0.66 \pm 0.28$ & $0.45 \pm 0.29$ & $0.26 \pm 0.04$ & $1.31 \pm 
0.19$  \\
 
6 & 0.1 & $0.39 \pm 0.28$ & $0.95 \pm 0.25$ & $0.18 \pm 0.03$ & $1.80 \pm 
0.26$  \\ \hline
\end{tabular}
\end{center}
\end{table}

\subsection{Spectroscopy in presence of \nm, \nt fluxes}

The spectroscopy can lead to quite different 
picture of suppression    
if one admits an  existence of the ``non electron" neutrino components.  
Now the effective flux measured by   
Kamiokande can be written as 
\be
F_{B}^{kam} \approx f_{B}\cdot \langle P_B(E)\rangle \cdot \langle F_B^0 
\rangle \cdot \xi ~, 
\label{bormod}
\ee
where 
\be 
\xi = 1 + r \frac{1 - \langle P_B \rangle }
{\langle P_B \rangle }~ .  
\label{xi}
\ee
The second term in the RH side is due 
to the neutral current 
scattering of non electron neutrinos, 
and $r \approx 0.15$ 
is the  ratio of the differential 
cross sections of the $\nm e -$   
and  the $\ne e - $ scattering.  
(In fact,  $r$ rather weakly depends on the energy of electrons for 
$E_e > 7$  MeV.)  

The averaged probability 
$\langle P_B \rangle$  
is unknown and in what follows we will perform the spectroscopy for 
different values  
$\langle P_B \rangle$  
(i.e. for different contributions of the 
\nm, \nt neutrinos to Kamiokande signal) 
keeping in mind that forthcoming 
experiments will be able to fix  
$\langle P_B \rangle$  along with the line 
described in sect. 2.2. \\

\noindent
{\it Step 1. Kamiokande.} 
The role of the second term in (\ref{xi}) increases with 
diminishing $\langle P_B \rangle$,  
so that the Kamiokande signal may be essentially 
due to \nm-,  \nt-  effect and in the limit $P \rightarrow 0$
one gets 
\be
F_B^{kam} \approx f_B \cdot \langle F_B^0 \rangle \cdot r~.  
\ee
This however implies that the original boron neutrino flux is 
much bigger than in the reference model: 
$
f_B \sim  R_{ e} r^{-1} \approx 2.8
$, 
( $R_{e} \equiv N_e^{kam}/ N_e^0$).  
As the consequence,  one expects 
large double ratio 
$(NC/CC)_{obs}/(NC/CC)_{SM} \gg 1$ 
in the SNO experiment. \\

\noindent
{\it Step 2. Homestake and  Kamiokande.} 
The electron neutrino flux is smaller than the flux 
measured by Kamiokande, $F^{kam}_B$.  
From (\ref{bormod}) we get the suppression factor for the 
electron neutrinos  can be estimated as 
\be
f_{B} \langle P_B \rangle \approx 
{R_{e}}{\xi}^{-1}, 
\label{eflux}
\ee
Correspondingly, the contribution of boron neutrinos 
to the $Ar$-production rate equals 
\be
Q_{Ar}^{B} \approx  
Q_{Ar,0}^{B} \xi^{-1} ~, 
\ee
where 
$Q_{Ar,0}^{B} (= 3.1)$ SNU is defined in (\ref{qar}).  
For $\langle P_B \rangle 
= 0.4,~ 0.3,~ 0.2$ we find  
$Q_{Ar}^{B} = 2.55,~ 2.30,~ 1.95 $ SNU respectively. 
The contribution of the boron neutrinos to 
$Ar$ - production decreases with 
$\langle P_B \rangle $,  
thus leaving the room for 
the beryllium neutrinos and other neutrinos of the 
intermediate energies. In particular,  at 
$\langle P_B \rangle \approx 0.4$ 
we get $Q_{Ar}^{B} \approx Q_{Ar}^{exp}$.  
For $\langle P_B \rangle  = 0.1$ the beryllium neutrinos may  
have (unsuppressed) RSSM - flux.\\ 

\noindent
{\it Step 3. Homestake,  Kamiokande and  Gallium.} 
 With decrease of $\langle P_B \rangle$ and therefore the increase of  
$f_{Be} P_{Be}$, the $pp$-neutrino flux should be suppressed to  
satisfy the gallium results.  The suppression factor equals  
\be
f_{pp} \langle P_{pp} \rangle = 
\frac{1}{Q^{SSM}_{pp}}\left[ Q_{Ge}^{exp} - 
Q_{Ge}^{Be} - 
Q_{Ge}^{B} - 
Q_{Ge}^{pep} - 
Q_{Ge}^{NO} 
\right]~. 
\label{fppw}
\ee
For completely suppressed beryllium flux, the Eq. (\ref{fppw}) 
reproduces result (\ref{fpp}). 
For unsuppressed beryllium flux we get 
\be
f_{pp} \langle P_{pp} \rangle \leq 0.40 \pm 0.11 ~. 
\ee
The results of spectroscopy for different values of 
$\langle P_B \rangle$ are shown in Table 2. 
With decrease of $\langle P_B \rangle $,   
the fluxes of intermediate energies,  
and in particular,  beryllium neutrino flux, allowed by data 
increase, whereas the suppression of the $pp$ - neutrino flux becomes 
stronger. Strong decrease of 
$\langle P_B \rangle $,  
implies big original boron neutrino flux. 
$$
f_B = \frac{R_{e}}{\langle P_B \rangle ~ \xi }~ . 
$$ 
The energy independent suppression is 
disfavored: If {\it e.g.}  $\langle P_B \rangle \sim 0.3$, then  
$f\cdot P = const$ is out of the data for more than $2 \sigma$ .

\subsection{Separate determination of $f_i$ and $P$}

As we discussed previously, further precise measurements of 
the distortion as 
well as the effects of \nm and \nt will give a possibility to determine 
$f_i$ and $\langle P_i \rangle$ separately. 
With present data one can get only  some limits 
for   $f_i$. 

1. Since $P \leq 1$,  there is the  lower bound  $f_i \geq f_i P_i $ . 
In particular,  from Kamiokande data we get 
\be
f_B > 0.3 ~ ~~~~ (2\sigma) . 
\ee
(Note that at $P = 1$ 
the fluxes of \nm and \nt are absent) . 

2. The upper bound on $f_B$ can be obtained in assumption 
that  the electron neutrinos convert into active neutrinos. For large $f_i$, 
the contribution from \nm and \nt alone can explain the data. 
Therefore $f_B < R_{\nu e} ~r^{-1} \approx 3.6$. Stronger  upper 
bound, $f_B < 2.8$ ($2 \sigma$),  
can be obtained,  if one takes into account also the 
Homestake result and suggests that the 
Argon is produced mainly by boron neutrinos.  
The limits can be further strengthen 
in the context of certain solution to 
the problem (see sect. 3).

3. The bounds on  $pp$- neutrinos can be obtained 
from the solar luminosity normalization condition 
(in assumption of thermal equilibrium of the Sun)    
\cite{spi,dar,bakr}:
\be
\sum_{i} \left(\frac{Q}{2} -  E_i \right) F_i = L_{\odot}  
\ee  
(here $E_i$ is the average energy of $i$-neutrino component, 
and $Q$ is the the energy release in the hydrogen cycle), 
and from 
the nuclear condition \cite{bakr}: 
\be
F_{pp} + F_{pep} \leq  F_{Be} + F_{B} ~~. 
\label{nucl}
\ee
This gives \cite{bakr} 
$$
0.5 < f_{pp} <  1.1 . 
$$  
The lower bound follows from (\ref{nucl}).  
The restriction is 
even more tight if one admits that $pp$-flux 
dominates over other fluxes. 
This gives essentially $f_{pp} = 1.00 \pm 0.05$ ,  
and consequently,  
$\langle P(E)\rangle  \approx 0.9 \pm 0.11$.  
To improve this number one should  continue the 
Gallium experiments and perform new  experiments like HELLAZ.   

4. The  $f_{Be}$ 
is restricted  very weakly. The model independent upper bound 
follows from the solar luminosity normalization condition: 
$f_{Be} < 6.35 $ \cite{bakr}.

\section{Implications}

Let us confront the  suppression profiles 
obtained from solar neutrino spectroscopy (Table 1.) with 
energy dependencies of different effects.

\subsection{Astrophysical solutions}

In this case $P_i (E) = 1$.  Majority of solutions  is  
based on  (or effectively reduced to) diminishing of 
the central temperature of the Sun.   
This gives  the  suppression profile with 
$f_{pp}: f_{Be}: f_{B}$ = $(T/T^{SSM})^{-1.2}$: 
$(T/T^{SSM})^{8 - 11}$:$(T/T^{SSM})^{18 - 25}$ = 
$1.05 : 0.7 :0.4$  
which should be compared with the profile \#1 in Table I.   
The intermediate energies are suppressed weaker than the high 
energies and this is the basis of statements  
that the astrophysical solutions are very strongly disfavored.  
To explain the data one needs more sophisticated  and more 
selective modification of the solar  model (see {\it e.g.} \cite{hax}). 
However  helioseismology gives very strong 
restrictions: 
The modifications which are consistent with helioseismology 
give only small changes of neutrino fluxes \cite{richa,vauc}.  
These aspects  have been   discussed lively in the talks  
by John Bahcall \cite{bahh} and  Arnon Dar 
\cite{darh}. 




\subsection{Vacuum oscillations solution}
 
Typical energy dependence of the survival 
probability \cite{grib}  
fits reasonably  well the profiles \#4,5 in the Table I.   
Rather big contribution from
\nm and \nt is  inferred. Basic features of the solution are 
\cite{bero} - \cite{calab} : 

(i)  For $\Delta m^2 > 3 \cdot 10^{-11}$ eV$^2$ 
the  boron neutrinos ($E > 7$ MeV) are  in the first high energy 
minimum of the suppression curve, $P(E)$.  The slope parameter, $s_e$,  
can be  bigger than in the MSW solution. 

(ii) The beryllium neutrinos are in the  
rapidly oscillating part  of the $P(E)$,  
so that one expects an appreciable 
{\it seasonal  variations} of the $Be$-neutrino flux due to annual 
change of distance 
between the Sun and the Earth.  
The strongest suppression is 
$P_{Be, min} = 1 - \sin^2 2\theta $. 
Large contribution of \nm and \nt to 
the Kamiokande signal resolves the 
problem of negative $f_{Be} P_{Be}$.  

(iii)  The $pp$-neutrino flux is in the 
region of averaged oscillations, where 
$P \sim  1 - 0.5 \sin^2 2\theta$. 
Lower experimental 
value of $Q_{Ge}^{exp}$ makes the fit better.  
It is easy to reach the 
inequality $P_{pp} > P_B > B_{Be}$ implied by spectroscopy. 
However,  there is an obvious relation 
between maximal suppression of the $Be$-line and suppression of 
$pp$-neutrinos: 
$P_{Be, min} = 2 P_{pp} -1$, 
and due to this the best fit configurations  
are  not realized for $f_i = 1$.  
The best fit ($\chi^2 \sim 2$  for 2 d.o.f.) corresponds to 
$\Delta m^2 = 0.6 \cdot 10^{-10}~ {\rm eV}^2,
~ \sin^2 2\theta = 0.9$  (for $f_i = 1$). 
Such a large  mixing  is disfavored by the 
data from SN1987A \cite{ssb}. 
Good  fit can be obtained 
also for moderate ($\sim 0.5$) suppression of the 
$Be$-line and $\sim 0.6$ suppression of 
the $pp$-neutrino flux. 

The fit becomes better for  values of 
$f_B$ bigger than 1 ~\cite{bero}. In this case    
the contribution of non-electron neutrinos 
to Kamiokande is large.  On the contrary, 
with diminishing $f_B$ 
a suppression of the boron neutrino flux due to  
oscillations should be  weaker. Therefore 
for fixed values of $\Delta m^2$ 
the allowed regions of parameters  
shift to smaller $\sin^2 2\theta$ ~\cite{bero,krpe1}.  
In particular, for 
$f_B = 0.7$,  
the region is  at  
$\sin^2 2\theta < 0.7$,  
thus satisfying the  bound from SN87A. 
For  $f_B \sim  0.4$  mixing can be as small as   
$\sin^2 2\theta < 0.5 - 0.6$. 

For  $f_B \sim   0.5$ the allowed region appears at 
$\Delta m^2 \sim 5 \cdot 10^{-12}$ eV$^2$~\cite{krpe2}   
which corresponds to  
the $Be$-neutrino line in the first high energy minimum of $P(E)$,  
the $pp$-neutrinos are in the first maximum of the $P$ and 
high energy part 
of the boron neutrino spectrum is out of suppression pit. 
No appreciable time variations are expected. 
Distortion of the $pp$-neutrino spectrum 
is the signature of this solution~
\cite{krpe2}. 

Depending on neutrino parameters as well as on  
$f_B$, $f_{Be}$ ...  one can get  variety 
of distortions of the boron neutrino energy 
spectrum~\cite{krpe1}.\\

\subsection{Resonance flavor conversion} 

{\it Small mixing solution}  \cite{msw}   
can precisely reproduce the desired energy dependence -- the 
profiles \#3, 4 from the Table 1   \cite{hala1, krsm,bfl,hala2}. In the 
region of small mixing angles  ($\sin^2 2\theta < 10^{-2}$) one has 
\begin{equation} 
P_{pp} \sim 1, \ \ \  P_{Be} \sim 0, \ \ \  
P_B \sim exp(-E_{na}/E), 
\label{eq:small}
\end{equation}
where $E_{na} = \Delta m^2 (\rho/\dot{\rho}) \sin^2 2\theta$   
and $\rho$ is the density profile. 

Let us sketch main properties of the  solution.  

(i) The dependence of the slope parameter on 
mixing angle  for 
$\Delta m^2 =6  \cdot 10^{-6}$ eV$^2$  is shown in the Table 2. 
The best fit slope corresponds to  
$\sin^2 2\theta = 4~10^{-3}$. 
At $2\sigma$ level the values $\sin^2 2\theta \sim 10^{-2}$  are  allowed.  

(ii) Additional contribution 
to Kamiokande, $\Delta f_B \approx 0.09$,  
follows from  scattering of the converted 
$\nu_{\mu}$ ($\nu_{\tau}$) 
on electrons. This solves the negative 
$f_{Be} P_{Be}$ problem.  

(iii) The $pp$-neutrino flux can be suppressed,  
if needed,  by diminishing $\Delta m^2$. In this case the 
high energy part of the spectrum will be in the 
suppression pit, and  one expects  distortion 
of spectrum of the $pp$-neutrinos.\\

With diminishing $f_B$ the  suppression due 
to conversion  should be   weakened,  and 
therefore  $\sin^2 2\theta$  decreases according to 
(\ref{eq:small}) ~\cite{krsm,bfl}. 
At  
$\Delta m^2 =6  \cdot 10^{-6}$ eV$^2$ 
the best fit of the data   
corresponds to the pairs of parameters~\cite{krsm}:  
$(f_B ,~ \sin^2 2\theta)$ = 
$(0.4,~ 1.0 \cdot 10^{-3})$,  
$(0.75,~ 4.3 \cdot 10^{-3})$, 
$(1.0, ~ 6.2 \cdot 10^{-3})$, 
$(1.5,~ 9 \cdot 10^{-3})$, 
$(2.0,~ 10^{-2})$ .  
The  decrease of $f_{Be}$ 
gives an additional small shift of the allowed region  
to  smaller values of $\sin^2 2\theta$.  
A consistent description 
of the data has been found for~\cite{krsm}
$$
f_B \sim 0.4 - 2.5 .
$$
Other fluxes are restricted rather weakly. 
At $2\sigma$ level: 
$f_{Be}  < 6.35 $ ,  and $f_{pp} = 0.55 - 1.08$
\cite{bakr}.\\ 

\noindent
For  {\it very small mixing solution}:  
$f_B \sim 0.5$, $\sin^2 2\theta \sim 10^{-3}$,   all  
the effects of the conversion become very small
in the high energy part of the boron neutrino spectrum 
($E > 5 - 6$ MeV). In particular,  
a distortion of the energy spectrum disappears, and the 
ratio  
$(CC/NC)^{exp}/(CC/NC)^{th}$ approaches 1. 
Thus studying just this 
part of spectrum, it will 
be difficult to identify the solution (e.g., to distinguish 
the conversion and the astrophysical effects) \cite{krsm} .\\ 
\begin{table}[t]
\caption{ 
The dependence of the slope parameter on mixing angle 
for $\Delta m^2 = 6 \cdot 10^{-6}$ eV$^{2}$. 
\label{tab:slope}
}
\vspace{0.4cm}
\begin{center}
\begin{tabular}{|c|c|c|c|c|c|}
\hline
$\sin^2 2\theta \times 10^{3}$  & 2     & 4    & 6    & 8     & 1.0 \\ 
\hline
$s_e$,  MeV$^{-1}$              & 0.017 & 0.034 & 0.046 & 0.055  & 0.065 \\ 
\hline
\end{tabular}
\end{center}
\end{table}

\noindent
{\it Large mixing MSW}  solution.   
The energy dependence of the effect gives reasonable approximation to 
the profiles \#4,5.   

(i) Boron neutrinos are in the bottom of the suppression pit, 
the slope parameter is very small  and  has negative sign. 
The day - night effect can be observed. 

(ii) Due to   
contribution from \nm and/or \nt leads the   inequality 
$P_{pp} > P_{B} > P_{Be}$ can be realized. 

(iii) For $pp$-neutrinos:  
$P \leq P^{vac} = 1 - \sin^2 2\theta /2$. 
Therefore small experimental  values 
of $Q_{Ge}^{exp}$  lead to better fit. 
  
The range of  neutrino parameters  is 
$
\Delta m^2 = (6 \cdot 10^{-6} - 10^{-4})~ {\rm eV^2}, \ \ \ 
\sin^2 2\theta = 0.65 - 0.85.
$  
With increase of $f_B$ a  suppression of the boron 
neutrino flux should increase and  the  fit of 
the data becomes better~\cite{krsm}. 
Now the Kamiokande signal can be explained 
essentially by NC effect and  mixing angle can be relatively small. 
Beryllium  neutrino flux is sufficiently suppressed and  
suppression of the $pp$-neutrinos is rather weak. 
For $f_B \sim 2 $,   values  
$\sin^2 2\theta = 0.2 - 0.3$ become allowed.\\ 

\noindent
Recently the effect of possible {\it density fluctuations} 
on the resonance 
conversion has been estimated \cite{fluc}. 
The fluctuations could be related to the gravity modes of oscillations 
of the Sun.  The 
effect leads to dumping of the neutrino conversion: a steady  
depart  of  neutrino state from the coherent mixture. 
As the result the probability of conversion 
approaches  1/2,  if the time of evolution is enough. 
In the case of varying average density the effects is mainly 
collected in the region of resonance layer. 
Small $\Delta \rho / \rho$ 
leads to  the biggest effect,  when 
the vacuum mixing angle is small {\it i.e.}  
the resonance occurs in 
the central parts of the Sun, 
and correspondingly $ P(E)$  is modified most strongly 
near the adiabatic edge. It can reach 
$\Delta P \sim 0.1 $ for  $\Delta \rho / \rho \sim 2\%$.  
The effect can be important for 
$^7 Be$- as well as $pp$- neutrinos. 
Effectively the adiabatic edge is shifted to higher $E$ which is 
equivalent to diminishing  $\Delta m^2$. 
Appreciable shift of the allowed region of neutrino parameters 
implies   $\Delta \rho / \rho \sim 5\%$. 

Note however,  the density fluctuations with typical scale 
$L\sim 10^2$ km and $\Delta \rho/ \rho \sim 2 - 5 \%$ 
may contradict helioseismological data.\\  

\noindent
{\it Conversion into 
sterile neutrino} 
$\ne \rightarrow S$. This solution differs from 
the flavor case in two points: 
(i). Effective matter density which determines the refraction effect 
is now $N = N_e - N_N/2$, {\it i.e.}, 
it depends also on density of neutrons. 
Since the concentration on neutrons is not  big even in the center of 
the Sun the effect is also quite small. (ii). There is 
no  contribution from $S$ in the Kamiokande detector,   
and the fit of the data is worser than in the flavor case. 
Therefore confirmation of negative value of 
$f_{Be} P_{Be}$  will disfavor this solution.   

The regions of neutrino parameters are approximately the same 
as in flavor case if one restricts the 
original boron neutrino flux by $f_B \leq  2$.\\

\noindent
{\it Three neutrino mixing}.  The analysis of data in terms 
of two neutrino mixing is quite realistic, since in the most 
interesting cases (simultaneous solution of the solar 
and hot dark matter problems, 
or solar and atmospheric neutrino problems)
the third neutrino has a large mass, so that 
its $\Delta m^2$ is beyond the solar resonance triangle region  and its 
mixing to the electron neutrino is rather small. This reduces the 
three neutrino task to the case of two neutrino mixing. 
However, there is an  interesting example, where third neutrino could  
influence solutions of the solar neutrino problem. 
It was considered previously~\cite{shi,asm,josh}
 and reanalyzed recently in~\cite{ema,lisi}. The third neutrino is in 
the region 
of a solution of the atmospheric neutrino problem: 
$m_3 \sim 0.1$ eV, and it has 
an appreciable admixture to the 
electron neutrino state:    
$\nu_e = \cos \phi ~ \nu' + \sin \phi ~ \nu_3 ,  
$ 
where 
$
\nu' = \cos \theta ~ \nu_1 + \sin \theta ~ \nu_2
$
and $\phi$ is not small. 
The third neutrino $\nu_3$ ``decouples" 
from the system  
(as far as we deal with the Sun) and 
its effect is reduced just to the averaged vacuum oscillations. 
In turn,  $\nu'$  converts resonantly into its   
orthogonal state. 
Therefore the survival probability can be written as~\cite{asm} 
\begin{equation}
P = \cos^4 \phi ~ P_2 + \sin^4 \phi, 
\label{eq:prob}
\end{equation}
where $P_2$ is the two neutrino survival probability. 
Additional regions of the neutrino parameters 
$\Delta m^2 = (10^{-5} - 10^{-6})$ eV$^2$ and 
$\sin^2 2\theta = 3\cdot 10^{-4} - 3\cdot 10^{-3}$ are allowed for 
$\cos^4 \phi \sim 0.5 - 0.7$. Both $pp$- and $Be$- neutrinos 
can be outside the $2\nu$ - suppression pit~\cite{asm}, 
where $P_2 \approx 1$ 
and according to (\ref{eq:prob})  the suppression factor for them 
is ($\cos^4 \phi + \sin^4 \phi$). This allows one to get 
about 1/2 suppression of the gallium production rate, 
and to reconcile  the 
Homestake and  Kamiokande results at $2\sigma$ level. 
Moreover,  now the boron neutrinos can be on the adiabatic edge of the 
suppression pit 
because of  distortion of the spectrum is weakened by 
factor $\cos^4 \phi$ and the slope parameter equals 
$s \approx \cos^4 \phi \cdot s_0$, 
($s_0$ is the slope in the 2$\nu$ case). The SK will certainly be able 
to check this scenario. 

A number of new possibilities appears, if both $\Delta m^2$ 
are in the resonance triangle of the Sun \cite{asm}, or 
if one of the resonances is in the resonance 
triangle whereas another one 
is in the region of ``just-so" solutions.

\subsection{Resonance spin-flavor precession}

As is well known the RSFP \cite{rsfp} leads to suppression factor 
which can 
perfectly reproduce the configuration 
of Fig. 2a \cite{last}. 
and which is very 
similar to that of small mixing MSW solution.  
There are however three important differences.  

\noindent
(i){\it  Asymptotic} at $E \rightarrow \infty$:  
\be
P \rightarrow 
\left\{ 
\begin{array}{ll}
1                              & {\rm for~~ the~~ MSW} \\ 
P_0(B, \Delta m^2, \theta ...) & {\rm for~~ the~~ RSFP}
\end{array}
\right .  ~. 
\ee
In the case of the RSFP the asymptotic value 
can be any number  in the interval $0 \leq P_0 \leq 1$ depending on the 
magnetic field profile, possible 
twist of the field, neutrino magnetic moment etc..  
If the $^7 Be$-line is strongly suppressed then 
the boron neutrino flux ($E > 7$ MeV) can be near the 
asymptotic region where the  slope parameter is  rather small. 

\noindent
(ii). {\it Correlation} of NC/CC and distortion. 
In contrast to the  MSW  2$\nu$-solution a distortion of 
spectrum is not correlated with absolute value of the probability. 
Strong suppression can be accompanied by weak distortion. 
One may observe large anomalous ratio 
NC/CC  in SNO and in the same time weak distortion 
of energy spectrum of the boron neutrinos. 
Note however that in the case of $3\nu$-resonance 
flavor conversion   the correlation can also be lost. 

\noindent
(iii). {\it Time variations}. There are strong time variations 
of the magnetic phenomena at the surface of the Sun 
and therefore it is difficult to expect that magnetic 
field profile is constant on the way of neutrinos 
inside the Sun or  that  changes are such that the 
integral effect is always the same. Therefore  
time variations are  generic consequences of  
this solution. Although a relation 
of the neutrino fluxes with surface activity can be rather complicated  
(neither simple correlations nor anticorrelations). 

It is  assumed that the  field which 
influences  neutrino propagation is the toroidal one. 
It has different   polarities in the 
northern and southern 
semispheres of the Sun,  so that in the equatorial plane 
the field has zero strength. The size of this 
equatorial gap is about  $5 - 7 ^0$ and 
one predicts seasonal variations of the neutrino flux 
due to presence of the gap and inclination of the Earth 
orbit with respect to the equatorial plane \cite{vvo}. 
No gap effect has been found by Kamiokande \cite{SK}.   

Reality of time variations of the solar neutrino fluxes  
is still open question. 
An  analysis shows that all experimental data 
are statistically compatible with constant neutrino fluxes
\cite{lissia}. However  even strong time 
variations, and in particular anti correlations with  
solar activity,  are not excluded. 

Suggested anticorrelations 
weakened during last 5 years. 
According to the analysis \cite{stan} 
the confidence level of the anticorrelation with 
sunspot number (as well as with some other 
characteristics of solar activity) 
declined very quickly since 1990 to 1994 from 0.9997 to 0.9.

In \cite{oak} the anticorrelations of the Homestake data with 
{\it surface magnetic flux} 
have been studied. Strong anticorrelation   
(probability $ \sim  99\%$) is found for the magnetic flux within central 
band 
($\pm 5^0$ of solar latitude) that has been delayed by 0.3 - 1.4 years 
with respect to neutrino signal. 
Still the results are not conclusive and the question of whether this 
anticorrelations are statistical or physical in nature requires data that 
span several solar cycles. 

\section{Conclusion}

1. We performed the solar neutrino spectroscopy using  data 
from four experiments. 
The  suppression profiles  $f_i P$ as the  
functions of the neutrino energy are found. The results 
depend strongly on whether the  
\nm,  \nt components are present in the solar neutrino flux. \\

\noindent
2. The  data give some  indication 
(``negative" beryllium neutrino flux) of  
presence of the non-electronic neutrinos in the neutrino flux. 
This result will be checked  
by\\ 
(i)  more precise measurements of signal in SK;\\  
(ii) measurements of the CC events in SNO 
and comparison with SK data;\\
(iii) measurements of the (NC)/(CC) ratio by SNO.\\

\noindent
3. Present data agree within $1\sigma$ with zero slope parameter, 
although rather strong distortion of the boron neutrino 
spectrum is not excluded. 
Measurements of the slope parameter of 
the recoil electron energy spectrum by SK, and 
directly the slope of neutrino energy spectrum by SNO  
will allow one  to find independently $f_B$ and $\langle P_B \rangle$,  
that is  to restore original boron neutrino flux and to get 
within given solution of the solar 
neutrino problem the bounds on neutrino parameters. 
These measurements will  give certain  discrimination 
among solutions. \\

\noindent
4. If there is no \nm and \nt neutrino flux, 
then the data lead to  energy profile 
\# 1 from the Table 1  with very strong  
suppression  at the intermediate energies.\\  

\noindent
5. If \nm -,   \nt -  components exist, then depending on 
$\langle P_B \rangle$ 
one may get  rather diverse  
dependences of $f\cdot P$ on energy. 
With decrease of $\langle P_B \rangle$ the  
fluxes of the intermediate energies,  
and in particular, the  beryllium neutrino flux, allowed by data,  
increase, whereas the suppression of the $pp$ - 
neutrino flux becomes 
stronger. Strong decrease of $\langle P_B \rangle$ 
implies, however,  big original boron 
neutrino flux.\\ 

\noindent
6. What can we expect after SK and SNO? 
Suppose  the SK will not find find neither  distortion of the recoil 
electron spectrum nor  the day - night effect. This may be explained by: 
(i) very small mixing MSW, 
(ii) large mixing MSW,  (iii) RSFP, (iv) very small $\Delta m^2$ vacuum 
oscillation solution, 
(v) averaged vacuum oscillations,  and (vi) still we will 
discuss astrophysical possibilities. 

Further discrimination can be done by SNO. If strong anomaly in 
ratio NC/CC will be found, then  solutions 
(i), (iv) and (vi) will be excluded. If SNO will not find 
NC/CC anomaly, only large mixing MSW conversion 
to the active neutrinos will be excluded and  BOREXINO results 
will be decisive.  

If SK will find distortions,  
then one should  discriminate among 
small mixing MSW, vacuum oscillations and   RSFP.     
If SK will  find large day-night effect, it will be the proof of the 
large  mixing MSW solution. 




\end{document}